\documentstyle[psfig,longtable]{aipproc}

\begin{document}
\title{Extracting $m_s$ From Flavor Breaking in Hadronic $\tau$ Decays}

\author{K. Maltman$^*$ and J. Kambor$^{**}$}
\address{$^*$Dept. Mathematics and Statistics, York Univ.,
4700 Keele St., Toronto, ON Canada, and CSSM, Univ. of Adelaide, 
Adelaide, SA Australia\thanks{Supported by the Natural Sciences
and Research Engineering Council of Canada} }
\address{$^{**}$Institut f\"ur Theor. Physik, Univ. Z\"urich,
CH-8057 Z\"urich, Switzerland\thanks{Supported by
the Schweizerischer Nationalfonds and
the EEC-TMR program under Contract No. CT 98-0169}}
%\lefthead{LEFT head}
%\righthead{RIGHT head}
\maketitle

\begin{abstract}
New finite energy sum rules (FESR's) for extracting $m_s$
from hadronic $\tau$ decay data
are constructed which (1) significantly
reduce potential theoretical uncertainties present in
existing sum rule analyses
and (2) remove problems associated with both
the poor convergence of the OPE representation of the
longitudinal part of the $us$ vector and
axial vector correlators and the large statistical errors
in the $us$ spectral data above the $K^*$ region.
\end{abstract}

The ratio of the hadronic $\tau$ decay rate through the $f=ij=ud,us$ 
vector (V) or axial vector (A) current to the corresponding
electronic decay rate, 
$R^{V/A;ij}_\tau 
= \Gamma [\tau^- \rightarrow \nu_\tau
\, {\rm hadrons}_{V/A;ij}\, (\gamma)]/ \Gamma [\tau^- \rightarrow
\nu_\tau e^- {\bar \nu}_e (\gamma)]$,
can be written\cite{pichrev}:
\begin{eqnarray}
{\frac{R^{V/A;ij}_\tau}{\left[ \vert V_{ij}\vert^2 S_{EW}\right]}} &=&
12 \pi^2 \, \int^{1}_0\, dy \,
\left( 1-y\right)^2 
\left[ \left( 1 + 2y\right) 
\, \rho_{V/A;ij}^{(0+1)}(s)  - \, 2y\, \rho_{V/A;ij}^{(0)}(s) \right] \, 
\nonumber \\
& =& 6 \pi \, i\, \oint_{|y|=1} dy \,
\left( 1- y\right)^2 \left[ \left( 
1 + 2y\right)\, \Pi_{V/A;ij}^{(0+1)}(s)
- 2 y\, \Pi_{V/A;ij}^{(0)}(s) \right] \ ,
\label{kinematicfesr}\end{eqnarray}
with $y=s/m_\tau^2$, $V_{ij}$ the $f=ij$ CKM matrix element, $S_{EW}$ 
an electroweak correction, and
$\rho^{(J)}_{V/A;ij}(s)$ the spectral function
of $\Pi^{(J)}_{V/A;ij}(s)$,
$\Pi^{(J)}_{V/A;ij}(s)$ ($J=0,1$) being the spin $J$ part of
the $f=ij$ vector (V) or axial vector (A) correlator.
The second line follows from the first 
as a consequence of the general FESR relation,
valid for any $\Pi$ without kinematic singularities, and any $w(s)$ analytic
in the region of the contour,
$\int_{s_{th}}^{s_0}\, ds\, w(s){\rho (s)}={\frac{-1}{2\pi i}}
\oint_{\vert s\vert =s_0}\, ds\, w(s)\Pi (s)$. 
Experimental data thus allows access to the $ud$-$us$ spectral
difference, and hence to integrals of the corresponding
correlator difference, which, for large enough $s_0$, are dominated
by the $D=2$ term of the OPE representation, proportional to $m_s^2$.
For such $s_0$ one may, therefore, hope to extract $m_s$ in terms
of appropriately weighted integrals of the experimental spectral data.

A complicating factor in any attempt to determine $m_s$
based on this observation is the non-convergence of the
OPE representation of the longitudinal ($(J)=(0)$) integral
at scales $\leq m_\tau^2$~\cite{kmtauprob}.
The current inability to make an experimental longitudinal/transverse
separation above $s\sim 1$ GeV$^2$ thus, at present, precludes
a reliable analysis using sum rules with
significant longitudinal contributions.  
Recent analyses\cite{ALEPHgroup,pp99}, which
either work with the spectral data without making
a longitudinal subtraction, or attempt to place loose experimental
bounds on the longitudinal contribution, 
employ ``spectral weights'' (for the
transverse ($(0+1)$) case, defining $y=s/s_0$, 
these are $w_\tau^k(y)\equiv
[(1-y)^2(1+2y)](1-y)^k$).
In this paper we work with the ``transverse''
$(V+A)^{(0+1)}_{ud}-(V+A)^{(0+1)}_{us}$ difference,
and try to find alternate weight choices which improve
the reliability of the analysis.
The $D=2$ term in the corresponding OPE is
\begin{equation}
\left[ \Pi^{(0+1)}_{V+A;ud}-\Pi^{(0+1)}_{V+A;us}\right]_{D=2}=
{\frac{3}{2\pi^2}}{\frac{
m_s(Q^2)}{Q^2}}\left[ 1+{\frac{7}{3}} a
+19.9332 a^2+\cdots \right] 
\label{d2}\end{equation}
where $a=\alpha_s(Q^2)/\pi$.
Details of the treatment of $D=4,6,8$ contributions
may be found in Ref.~\cite{jkkmms}.

On the spectral side we employ the ALEPH $ud$ and
$us$ distributions\cite{ALEPHgroup,ALEPHud}.
The longitudinal subtraction is performed using sum rule 
methods~\cite{jkkmms};
for the weights employed in our analysis 
the integrated longitudinal
subtraction represents $<1\%$ of the integrated $D=2$
OPE $ud$-$us$ {\it difference}, 
making the impact of any uncertainties associated
with this procedure negligible.

The weights employed in the $ud$-$us$ FESR's of our analysis have been
chosen so as to reduce both theoretical and experimental difficulties.
On the experimental side, we seek to 
(1) de-emphasize $us$ spectral contributions from the region above
the $K^*$, since the ALEPH determination of 
$\rho_{V+A;us}$ has 
$\sim 20-30\%$ statistical
errors in this region~\cite{ALEPHgroup} and (2) reduce the
strong cancellation in the $ud$-$us$
difference, which otherwise greatly magnifies the impact of
experimental uncertainties.  (See Ref.~\cite{jkkmms} for
a detailed discussion of the second point.)
Weights which fall
more strongly with $s$ above $s\sim 1$ GeV$^2$
decrease the level of $ud$-$us$ cancellation and simultaneously
suppress high-$s$ contributions, decreasing
the impact of both the longitudinal
subtraction and experimental errors on $\rho_{V+A;us}$.

\noindent
\begin{figure} [htb]
\centering{\
\psfig{figure=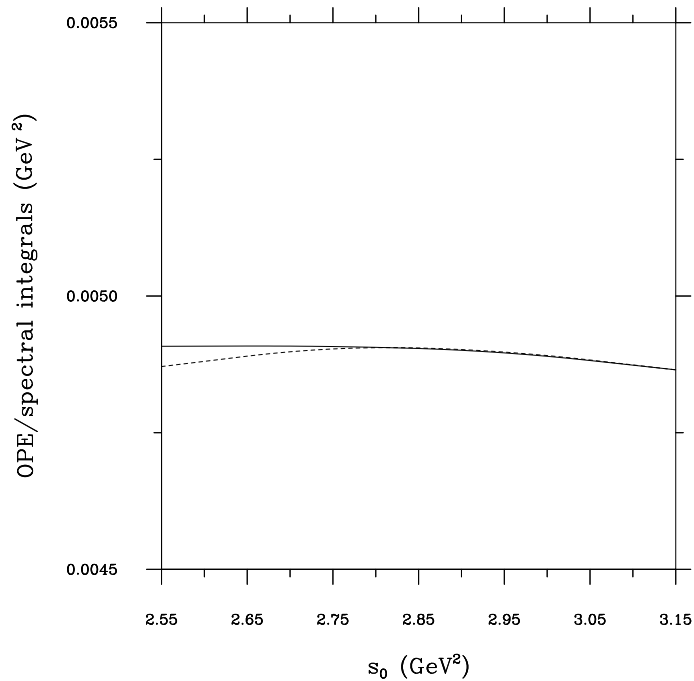,height=10.cm}}
\vskip .3in
\caption{The agreement between the OPE and hadronic sides of
the FESR corresponding to the weight, $w_{20}(y)$ for
$2.55\ {\rm GeV}^2\leq s_0\leq m_\tau^2$.
The solid line is the OPE side, using
the values of $m_s$ and the $D=8$ contribution
obtained in the fitting procedure
described in the text and, in more detail, in Ref.~[5].
The dashed line is the
hadronic side, obtained using the ALEPH spectral data from which
the longitudinal component has been subtracted as described
in Ref.~[5].}
\label{figtwo}
\end{figure}

On the theoretical side, the goal is to
control an important potential
theoretical systematic uncertainty.  For the weights
$w_\tau^k$, the contour improvement prescription~\cite{cipt}
is known to produce 
a significant improvement in the convergence of the
known terms of the integrated $D=2$ series.
The smallness of the last ($O(\alpha_s^2)$) known term, however, 
turns out to result from strong cancellations between
contributions from different regions of the circular part
of the contour, $\vert s\vert =s_0$~\cite{jkkmms}.  Since,
assuming continued geometric growth of the coefficients, 
similar cancellations do
not persist to higher orders~\cite{jkkmms}, an estimate 
of the truncation error based on the size of the $O(\alpha_s^2)$
term is unreliable.
We have constructed 3 alternate 
polynomial weights which both avoid such ``accidental''
cancellations and emphasize regions of the plane for 
which the convergence of the $D=2$ series is optimal.
The result is a very strong suppression of possible
higher order $D=2$ contributions~\cite{jkkmms}.
The explicit forms of the weights, as well as details of this
improvement, are given in Ref.~\cite{jkkmms}.

The results of our analysis are as follows.  
First, all 3 new weights yield consistent, and stable,
values of $m_s$ in the window $2.55\ {\rm GeV}^2<s_0<m_\tau^2$.  
An illustration of this fact is given, in Table 1, for
the weight $w_{10}(y)=[1-y]^4[1+y]^2[1+y^2][1+ y+y^2]=
1-y-y^2+2y^5-y^8-y^9+y^{10}$, where $y=s/s_0$, which is
favorable from a theoretical point of view because the
absence of $y^3$, $y^4$ terms removes $D=8, 10$ contributions
to the integrated OPE. 
Second, for our two other weights, which do not share this
property,
the $D=8$ contributions, which are determined
self-consistently, are also stable in this window.
These conditions are not satisfied for the FESR's based
on the spectral weights, $w_\tau^k$.
We choose, for our final analysis, that weight among the
three constructed above (called $w_{20}$ in Ref.~\cite{jkkmms})
which leads to the smallest fractional statistical error.
The match between the OPE and hadronic sides of the
corresponding FESR which results once $m_s$ and the
$D=8$ contribution have been optimized is shown in Figure 1,
and is clearly excellent.
Our final numerical result for $m_s$, 
in the $\overline{MS}$ scheme, is 
\begin{equation}
m_s(1\ {\rm GeV}^2)=158.6\pm 18.7\pm 16.3\pm 13.3\ {\rm MeV}\ ,
\end{equation}
which, using four-loop running, corresponds to
\begin{equation}
m_s(4\ {\rm GeV}^2)=115.1\pm 13.6\pm 11.8\pm 9.7\ {\rm MeV}\ .
\end{equation}
The first error is statistical, 
the second due to the uncertainty in $\vert V_{us}\vert$, 
and the third theoretical, with the latter dominated by
our estimate of the error associated with truncating
the $D=2$ series at $O(\alpha_s^2)$.  Improvements
in the accuracy of the $us$ spectral data, such as
will be possible at BaBar, will serve to significantly
reduce the first error and, simultaneously, allow
use of weights other than $w_{20}$ which produce reduced 
theoretical truncation errors.

\noindent
\begin{table}
\caption{The extracted value of $m_s (1\ {\rm GeV}^2)$ in MeV as
a function of $s_0$ for the weight $w_{10}$ having no $D=8,10$
contributions.}
\begin{tabular}{lccccc}
$s_0$ (GeV$^2$):&2.35&2.55&2.75&2.95&3.15 \\
$m_s (1\ {\rm GeV}^2)$ (MeV):&153.2&159.0&162.2&163.4&163.2 \\
\end{tabular}\label{table3}
\end{table}

\end{document}